\documentclass[aps,prd,twocolumn,superscriptaddress,preprintnumbers,floatfix,nofootinbib,notitlepage,showkeys,showpacs]{revtex4-1}

\usepackage[utf8x]{inputenc}

\usepackage{graphicx}
\usepackage{hyperref}
\usepackage{latexsym}
\usepackage{amsmath}
\usepackage{amssymb}
\usepackage{bbm}
\usepackage{ulem}
\usepackage{pdfsync}
\usepackage{epsfig}
\usepackage{epstopdf}
\usepackage{subfigure}
\usepackage{color}
\usepackage{comment}
\usepackage{slashed}

\newcommand{\be}{\begin{equation}}
\newcommand{\ee}{\end{equation}}
\newcommand{\bea}{\begin{equation}\begin{aligned}}
\newcommand{\eea}{\end{aligned}\end{equation}}

\def\lsim{\mathrel{\raise.3ex\hbox{$<$\kern-.75em\lower1ex\hbox{$\sim$}}}}
\def\gsim{\mathrel{\raise.3ex\hbox{$>$\kern-.75em\lower1ex\hbox{$\sim$}}}}

\begin{document}

\title{Primordial Black Hole Formation During Slow Reheating After Inflation}

\author{Bernard Carr}
\email{b.j.carr@qmul.ac.uk}
\affiliation{Astronomy Unit, Queen Mary University of London, \\ Mile End Road, London, E1 4NS, United Kingdom}
\author{Konstantinos Dimopoulos}
\email{konst.dimopoulos@lancaster.ac.uk}
\affiliation{Consortium for Fundamental Physics, Physics Department,Lancaster University, Lancaster LA1 4YB, United Kingdom}
\author{Charlotte Owen}
\email{c.owen@lancaster.ac.uk}
\affiliation{Consortium for Fundamental Physics, Physics Department,Lancaster University, Lancaster LA1 4YB, United Kingdom}
\author{Tommi Tenkanen}
\email{t.tenkanen@qmul.ac.uk}
\affiliation{Astronomy Unit, Queen Mary University of London, \\ Mile End Road, London, E1 4NS, United Kingdom}

\begin{abstract}
We study the formation of primordial black holes (PBHs) in the early Universe during a period of slow reheating after inflation.
We demonstrate how the PBH formation mechanism may change even before the end of the matter-dominated phase and calculate the expected PBH mass function. We find that there is a threshold for the variance of the density contrast, $\sigma_c \simeq 0.05$, below which the transition occurs even before reheating, with this having important consequences for the PBH mass function. We also show that there is a maximum cut-off for the PBH mass at around $100\,M_{\odot}$, below which the subdominant radiation bath affects PBH production, making the scenario particularly interesting for the recent LIGO observations of black hole mergers.
\end{abstract}

\maketitle

%%%%%%%%%%%%%%%%%%%%%%%%%%%%%%%%%%%%%%%%%%%%%%%%%%%%%%%%
% DOCUMENT
%%%%%%%%%%%%%%%%%%%%%%%%%%%%%%%%%%%%%%%%%%%%%%%%%%%%%%%%

%%%%%%%%%%%%%%%%%%%%%%%%%%%%%%%%%%%%%%%%%%%%%%%%%%%%%%%%
\section{Introduction}

Primordial black holes (PBHs) can form in the early Universe as a result of the huge compression during the Big Bang \cite{Hawking:1971aa}. A comparison of the cosmological density at time $t$ with the density of a black hole of mass  $M$ implies that they will have around the horizon mass at formation ($M \sim c^3t/G$).  Such black holes are the only ones which could be small enough for quantum effects to be important, those forming before $10^{-23}$s being smaller than $10^{15}$g and therefore evaporating by the current epoch \cite{Hawking:1974rv}. Larger ones might provide the dark matter (DM) or have other interesting astrophysical consequences.

PBHs can form through a variety of mechanisms but the most natural one - certainly the scenario first considered historically - is that they derive from  primordial inhomogeneities. An overdense region in the early Universe  can collapse to a black hole provided it is larger than the Jeans length at maximum expansion, which is $R_J \approx \sqrt{w} \, ct$ for an equation of state $p = w \rho c^2$ \cite{Carr:1974nx}. A simple heuristic argument then requires  the density fluctuation to exceed a critical value $\delta_c \approx w$ at horizon crossing \cite{Carr:1975qj}. Generally the expected fluctuation $\sigma(M)$ will be much smaller than this, so only the small fraction of regions on the tail of the fluctuation distribution  are expected to  form PBHs. In most scenarios one expects the fluctuations to be Gaussian, so
this fraction  is $\beta(M) \propto \exp (- w^2/\sigma(M)^2)$ and therefore exponentially suppressed \cite{Carr:1975qj}. 

The fact that the fraction of the Universe collapsing is tiny does not necessarily imply that the PBHs are unlikely to have formed since one also {\it requires} the collapse fraction to be small. This is because the ratio of the PBH density to the background density increases with the cosmic scale factor $a$ as $a^{3w} \propto t^{2w/(1+3w)}$ and one needs the PBHs to have less than the observed dark matter density today, $\Omega_{\rm PBH} < \Omega_{\rm DM} \approx 0.26$ \cite{Ade:2015xua}. If the early Universe is radiation-dominated (RD) ($w=1/3$), as applies during most epochs before matter-radiation equality ($t_{\rm eq} \sim 10^{12}$s), then one requires $\beta(M) < 10^{-9} \Omega_{\rm PBH}(M) (M/M_{\odot})^{1/2}$ \cite{Carr:1975qj}. Thus one both expects and requires the collapse fraction to be small. 

These arguments were first given more than 40 years ago and have subsequently been refined in many ways. For example, hydrodynamical calculations have been used to determine the critical value $\delta_c$ for more realistic initial density profiles in the collapsing region, incorporating the effects of pressure gradients \cite{1978SvA....22..129N}. Also the fluctuations which generate PBHs are initially much larger than the cosmological horizon and the precise definition of such an overdensity is gauge-dependent and non-unique. Modern treatments often use the curvature perturbation $\zeta$, which is a measure of the total energy perturbation \cite{Kopp:2010sh,Harada:2013epa}. More importantly, one can view PBH formation as a manifestation of critical phenomena, first discovered in a non-cosmological context \cite{Choptuik:1992jv}, which arise for a very general form for the initial density profile \cite{Niemeyer:1999ak}. In this case, there is still a critical threshold $\delta_c$ and one can determine this as a function of the equation of state parameter $w$ \cite{Musco:2004ak,Musco:2008hv,Musco:2012au}. Most of the PBHs still have the horizon mass at formation but their mass spectrum now extends down to much smaller masses. 

There has also been much interest in the form of the initial density fluctuations, with many authors  considering those expected in the inflationary scenario. Indeed PBH formation provides a unique test of inflation precisely because it probes the power spectrum on scales which are too small to observe directly. In this context, it is important to stress that the amplitude of the fluctuations on cosmological scales, as probed by the cosmic microwave background and large-scale structure observations, is only $\sigma \sim 10^{-5}$ \cite{Ade:2015lrj}. 
This means that 
 $\beta$ is too small to be interesting, at least  if one believes in its exponential dependence on $\sigma$. 

One way to circumvent this conclusion is to assume that $\sigma(M)$ is increased on small scales -- either because  the fluctuations have a blue spectrum \cite{Carr:1994ar,Carr:2017edp} or because there is some special feature in power-spectrum \cite{Bugaev:2008bi,Garcia-Bellido:2017mdw}. Also many authors have considered the effects of non-Gaussian fluctuations since these arise naturally in some inflationary scenarios \cite{Young:2013oia,Young:2015kda,Franciolini:2018vbk}.
However, even in this case $\beta$ has an exponential dependence on $\sigma$, so it still requires fine-tuning if the PBHs are to provide the dark matter \cite{Carr:2016drx} or generate the gravitational wave-bursts detected  by LIGO \cite{Bird:2016dcv}.

Another way of enhancing PBH formation -- and the main focus of the present paper -- is to assume that the pressure is reduced at some epoch. Because of the exponential dependence of $\beta$ upon $w$, even a small reduction in $w$ can boost $\beta$ enormously. For example, it has been argued that this may occur at the QCD phase transition \cite{Byrnes:2018clq}, when the horizon mass is around \mbox{$1\,M_{\odot}$},
and this could be relevant to the LIGO observations. 

More dramatically, it has been proposed that  the Universe may become effectively  pressureless for some early period. For example, this may occur due to its energy density being channelled into non-relativistic massive particles \cite{Khlopov:1980mg,Polnarev:1986bi,Carr:2017edp} or due to slow reheating after inflation \cite{Carr:1994ar,Alabidi:2013wtp,Suyama:2004mz,Suyama:2006sr,Hidalgo:2017dfp}. In the latter case, the field which is responsible for reheating the Universe typically oscillates about the minimum of its potential $V(\phi)$, with $w$ having an average value of zero.

During any pressureless (matter-dominated) early phase, $w$ becomes sufficiently small to remove the exponential suppression of $\beta$. Physically this is because the Jeans length  becomes so small that most overdense regions would be expected to collapse. However, in this case another factor comes into play: the region can only collapse into a PBH if it is sufficiently spherical. If it is aspherical or has angular momentum, it is likely to form a disc or fragment in some way (as happens to overdense regions generating galaxies after decoupling). 

This does not apply when the Jeans length is comparable to the cosmological horizon because the region does not collapse much between entering this horizon and forming a black hole. Also one expects overdense regions on the Gaussian tail to be spherically symmetric \cite{1970Ap......6..320D,1986ApJ...304...15B}. However, during a matter-dominated phase, one can show that the probability of a region being sufficiently spherically symmetric to form a PBH is $\beta \sim \sigma^5$ for sufficiently large $\sigma$ \cite{Khlopov:1980mg,Polnarev:1986bi,Harada:2016mhb, Kuhnel:2016exn}), although this expression must be modified if the region has spin \cite{Harada:2017fjm}. For $\sigma \sim 10^{-5}$, this is still too small to explain the dark matter or LIGO observations. Nevertheless, it is much larger than the Gaussian expression for $\beta$, so the variation in the power-spectrum required for PBHs is much more modest than in the standard radiation-dominated scenario.

In this paper, we consider scenarios in which the Universe is matter-dominated during an oscillatory phase after inflation due to slow reheating. We go beyond the usual calculation by allowing for the (initially subdominant) component of radiation generated by the gradual decay of the oscillating scalar field.  This means that the exponential suppression is initially unimportant, with lack of spherical symmetry and non-vanishing angular momentum opposing collapse. However, it  gradually becomes more significant as the value of $w$ increases and eventually there is a transition in which the exponential expression for $\beta$ falls below the $\sigma^5$ expression. This gives a natural upper cut-off in the PBH mass function, which is not included in the usual treatment of matter-dominated scenarios.

The paper is organized as follows. In Section \ref{reheating}, we demonstrate how the formation mechanism for PBHs slowly changes in the transition from the matter-dominated to radiation-dominated phases.
In Section \ref{sec:massfunction}, we calculate the characteristic PBH mass function resulting from this transition.  Finally, in Section \ref{conclusions}, we present our conclusions.

\section{Primordial black hole production during reheating}
\label{reheating}

We assume that reheating proceeds as follows: the scalar field $\phi$ responsible for reheating the Universe oscillates in a quadratic potential, $\rho_\phi = 3H^2M_{\rm P}^2 \propto a^{-3}$, so that the Universe is effectively matter-dominated from the end of inflation until the time when the scalar condensate decays into radiation. Here $\rho_\phi$ is the energy density of the field $\phi$, $H$ is the Hubble scale, $M_{\rm P}$ is the reduced Planck mass, and we use natural units with $\hbar =  c =1$ and \mbox{$8\pi G=M_{\rm P}^{-2}$}. 

Reheating ends when $\Gamma=H$, where $\Gamma$ is the decay rate of the scalar field into radiation. By using the Friedmann equation, one can express the reheat temperature as a function of the decay rate:
\be
\label{Treh}
T_{\rm reh} = \left(\frac{90}{\pi^2 g_*(T_{\rm reh})}\right)^{1/4}\sqrt{\Gamma M_{\rm P}} \, .
\ee
Here $g_*$ is the effective number of degrees of freedom in the radiation heat bath, which we assume forms and maintains its equilibrium state throughout reheating. Note that the reheat temperature is fully determined by $\Gamma$, with the duration of the reheating era depending only on this quantity. In this paper we do not specify the value of $\Gamma$ but use it as a free parameter for which exact value (or time-dependence) can be given once a concrete model has been specified. There are, however, model-independent upper and lower bounds on $\Gamma$: for instant reheating the upper bound on the inflationary scale requires $\Gamma \leq 
%H_*\simeq 
10^{14}$ GeV \cite{Ade:2015xua}, while retaining successful Big Bang Nucleosynthesis (BBN) requires $T_{\rm reh}\gtrsim 4$ MeV~\cite{Kawasaki:2000en,Hannestad:2004px,Ichikawa:2005vw,DeBernardis:2008zz}, giving a bound $\Gamma \gtrsim 10^{-37}$ GeV for the matter content of the Standard Model of particle physics with $g_*(4 \, {\rm MeV})=10.75$.

Before reheating, the subdominant radiation component had an energy density \cite{Kolb:1990vq}
\be
\label{rhorad}
\rho_{\rm r} = \frac{\Gamma}{4 H}\rho_\phi .
\ee
Using this, the effective equation of state parameter can be written as
\be
\label{weff}
w_{\rm eff}=\frac{p_\phi + p_{\rm r}}{\rho_\phi + \rho_{\rm r}} 
=w_r\frac{\Gamma}{4H}\left(1+\frac{\Gamma}{4H}\right)^{-1}
\simeq 
\frac{w_{\rm r}}{4}\frac{\Gamma}{H} ,
\ee
where $p_i$ and $\rho_i$ are, respectively, the pressure and energy density of the $\phi$ field and the subdominant radiation component, $w_{\rm r}=1/3$ and
we have taken $p_\phi \simeq 0$. For the last equality above we have assumed $\rho_{\rm r}\ll \rho_\phi$.

When $\Gamma\ll H$, the Universe is effectively matter-dominated (MD) and the fraction of the total energy density collapsing into PBHs of mass $M$ is usually taken to be \cite{Khlopov:1980mg,Polnarev:1986bi,Harada:2016mhb}
\be
\beta(M) \simeq
0.056\sigma(M)^5 \, ,
\label{kp}
\ee
where $\sigma(M)$ is the variance of density perturbations at the time the total mass inside the horizon is $M$. 
Recently Harada et al. \cite{Harada:2017fjm} have modified Eq.~(\ref{kp}) to allow for the effect of rotation in the collapsing region. The expression is unchanged for $0.005\lesssim \sigma(M) \lesssim 0.2$ but replaced by
\be
\label{betamat}
\beta(M) \simeq
2\times 10^{-6}f_q(q_c)\mathcal{I}^6\sigma(M)^2{\rm exp}\left(-0.147\frac{\mathcal{I}^{4/3}}{\sigma(M)^{2/3}}\right) 
\ee
 for $\sigma(M) \lesssim 0.005$.
Here 
$f_q(q_c)$ is the fraction of masses for which the dimensionless quadrupole moment $q$ is smaller than some critical value. Following Ref.~\cite{Harada:2017fjm}, we take this to be $q_c=\sqrt{2}$ and we also take $\mathcal{I}=1$ and $f_q\sim 10^{-5}$. 

Note that Refs.~\cite{Khlopov:1980mg,Polnarev:1986bi}
include an extra suppression factor
$\sigma^{3/2}$ in Eq. \eqref{kp} to account for inhomogeneity
effects and this would decrease the probability of PBH
formation. However, as discussed in Ref.~\cite{Harada:2016mhb}, there is some
uncertainty about this factor, so we neglect it here. 

In any case, as the ratio $\Gamma/H$ approaches unity, the  matter-dominance approximation breaks down and the Universe slowly enters the radiation-dominated era. For spherically symmetric regions, the fraction of the total energy density collapsing into PBHs of mass $M$ is then \cite{Carr:1975qj}
\bea
\label{betarad}
\beta(M) &=  \frac{2}{\sqrt{2\pi}\sigma(M)}\int_{\delta_c}^\infty {\rm d}\delta\, \exp\left(-\frac{\delta^2}{2\sigma(M)^2}\right) \\
&= {\rm Erfc}\left( 
\frac{\delta_c}{\sqrt{2}\sigma(M)} \right) \,,
\eea
where ${\rm Erfc}$ is the complementary error function. A precise expression for $\delta_c$ has been given by Harada et al. \cite{Harada:2013epa}:
\be
\delta_c = \frac{3(1+w)}{5+3w}\sin ^2 \left( \frac{ \pi \sqrt{w}}{1+3w} \right)\, .
\ee
This applies in the comoving gauge, which is the appropriate choice if one wishes to compare to Eqs. (\ref{kp}) and (\ref{betamat}). For $w \ll 1$, this just gives $\delta_c \approx (3\pi^2/5) w \approx 6w$.

Even though the radiation component remains subdominant until reheating, $\rho_{\rm r}\ll \rho_{\phi}$, the transition between the two production regimes occurs when the probabilities for PBH formation in matter-dominated and radiation-dominated eras coincide. 
Combining Eqs.~\eqref{weff} and \eqref{betarad} we obtain
\be
\label{GoverH}
\frac{\Gamma}{H} \simeq \frac{2\sqrt{2}\sigma}{3w_{\rm r}}{\rm Erfc}^{-1}
\left( \beta(\sigma)\right)\,,
\ee
where ${\rm Erfc}^{-1}$ is the inverse complementary error function and
we have used 
$\delta_c = 6 w_{\rm eff}$ with $w_{\rm eff}$ given by Eq.~\eqref{weff}. 
In the above, $\beta$ is given by either Eq.~\eqref{kp}
or Eq.~\eqref{betamat}, 
within their range of validity.

\begin{figure}
%%%%
\begin{center}
%%%%
\includegraphics[width=.45\textwidth]{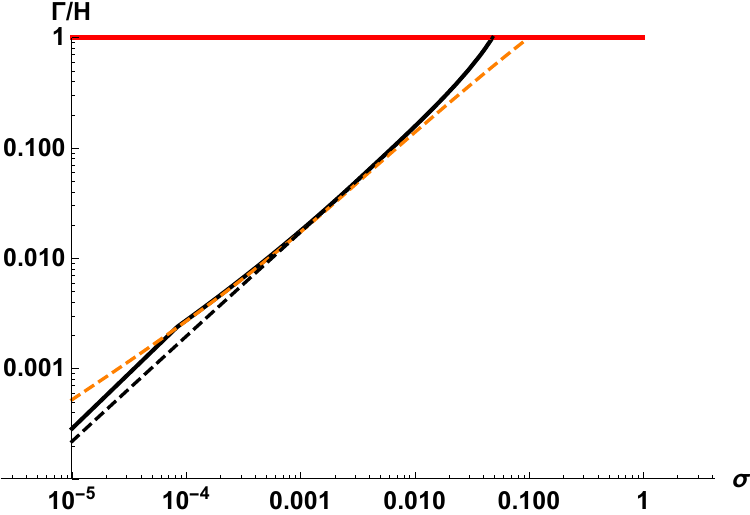}\\
%%%%
\caption{The value of $\Gamma/H$ at 
the transition between the different PBH production 
mechanisms 
for a given value of the variance of the density contrast $\sigma$ (black solid line). The orange dashed line shows a fit to the numerical result, obtained by putting
$\delta_c = 6 w_{\rm eff}$. The black dashed line shows the result 
when the effect of angular momentum is 
omitted. The red horizontal line 
corresponds to $\Gamma=H$, 
above which the Universe remains effectively matter-dominated until reheating.}
%%%%
\label{GammaH}
%%%%
\end{center}
%%%%
\end{figure}

We find that the values of
$\Gamma/H$ 
determined by
Eq.~\eqref{kp} and Eq.~\eqref{betamat} are roughly the same for 
\mbox{$5\times 10^{-5}\leq\sigma\leq 0.01$}.
This coincidence does not describe the transition accurately for $\sigma < 5\times 10^{-5}$ or $\sigma > 0.01$ but the error is at most a factor of two. The exact result can be found numerically, as shown in Fig.~\ref{GammaH}.
We find that $\Gamma/H = 1$ for $\sigma_c \simeq 0.05$, i.e. for perturbations larger than $\sigma_c$ the transition between production mechanisms never occurs before reheating is completed and so Eq.~\eqref{betamat} applies throughout the period of slow reheating. For $\sigma < \sigma_c$ the production mechanism changes before reheating. We show below that this has important consequences for the PBH mass function.

\begin{figure}
%%%%
\begin{center}
%%%%
\includegraphics[width=.45\textwidth]{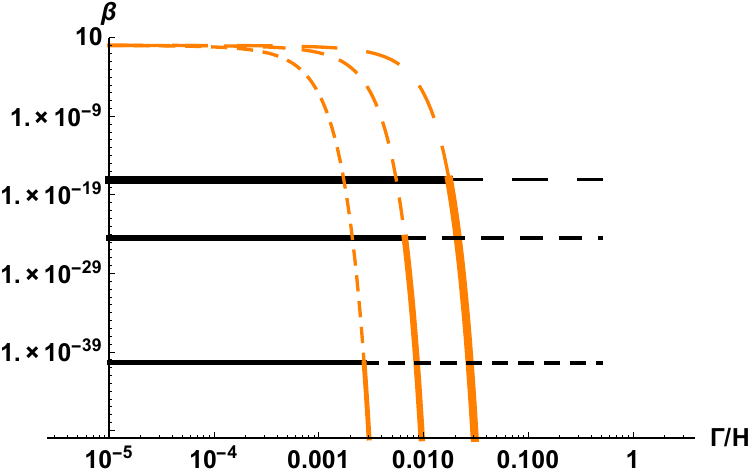}\\
%%%%
\caption{The fraction of the total energy density collapsing into PBHs, $\beta$, versus the decay rate of the field responsible for reheating, $\Gamma/H$, for 
matter-dominated (horizontal black lines) and 
radiation-dominated  (orange curves) scenarios. The solid lines show the 
fraction for 
$\sigma=10^{-3},10^{-3.5},10^{-4}$ (from top to bottom), while 
the dashed lines are 
continuations of the MD and RD results.}
\label{betaGamma}
%%%%
\end{center}
%%%%
\end{figure}

\begin{figure}
%%%%
\begin{center}
%%%%
\includegraphics[width=.45\textwidth]{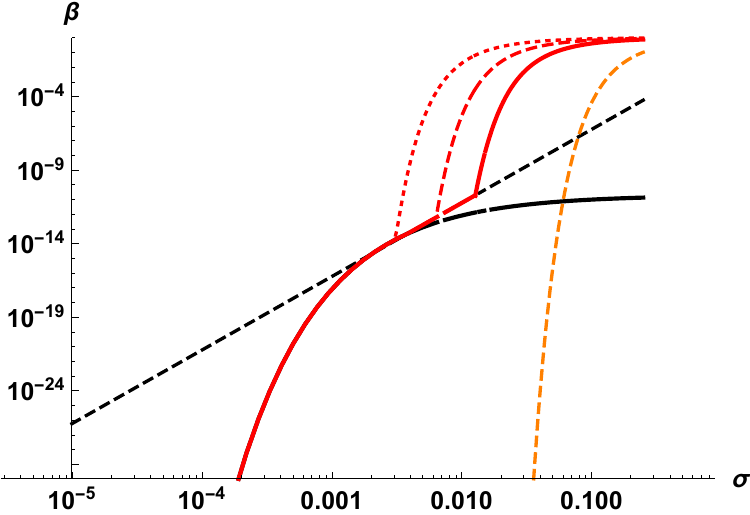}\\
%%%%
\caption{The fraction of the total energy density collapsing into PBHs, $\beta$, as a function of the variance of the density contrast, $\sigma$, for $\Gamma/H=0.05$ (red dotted line), $\Gamma/H=0.1$ (red dashed line) and $\Gamma/H=0.2$ (red solid line). The black solid (dashed) line is 
the matter-dominated case with (without) spin effects
and the orange dashed line is 
the usual radiation-dominated case with $w=1/3$.}
%%%%
\label{betaSigma}
%%%%
\end{center}
%%%%
\end{figure}

The results are shown in Figs. \ref{GammaH}, \ref{betaGamma} and \ref{betaSigma}. The probability of an overdense regions collapsing to a PBH is the product of three factors: $(i)$ the probability of its being larger than the Jeans mass at maximum expansion; $(ii)$ the probability of its being sufficiently spherically symmetric; and $(iii)$ the probability of its spinning sufficiently slowly. The first probability is close to unity in the matter-dominated era; the second probability is close to unity in the radiation era because regions on the Gaussian tail (as required) are likely to be nearly spherical. In this paper we allow a smooth transition in probability $(i)$, whereas we have a discontinuous transition in probability $(ii)$.

%%%%

%%%%

%%%%

\begin{figure}
%%%%
\begin{center}
%%%%
\includegraphics[width=.45\textwidth]{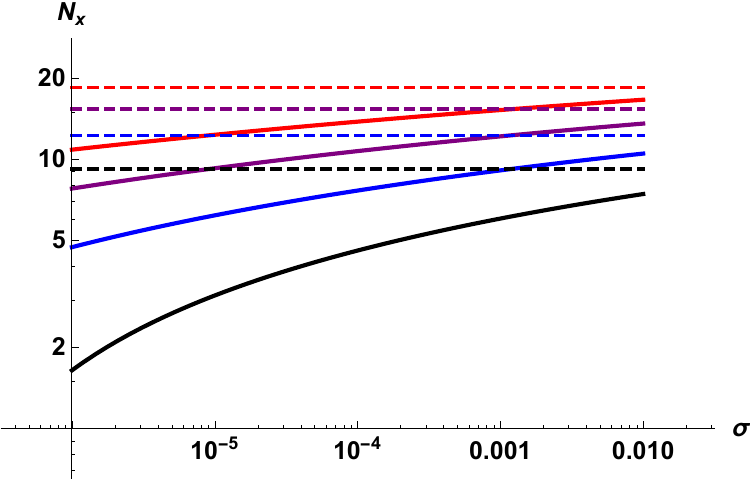}\\
%%%%
\includegraphics[width=.45\textwidth]{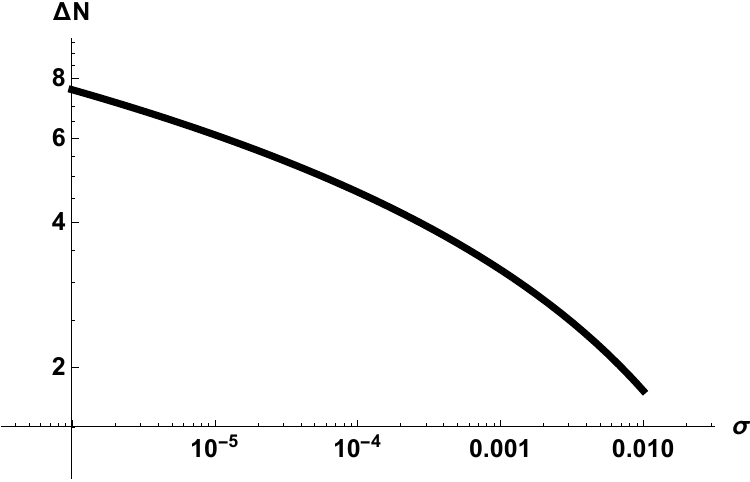}\\
%%%%
\caption{{\it Upper panel}: The solid lines show the number of e-folds, $N_{\rm x}$, after which
PBH production is  suppressed as a function of the primordial 
perturbation $\sigma$ for $\Gamma/H_*=10^{-12},10^{-10},10^{-8},10^{-6}$ (top to bottom). 
The dashed lines show the standard result $N_{\rm reh}$ for the same $\Gamma/H_*$ as above.
{\it Lower panel}: The change in the number of e-folds of reheating, 
$\Delta N = N_{\mathrm{reh}} - N_{\mathrm{x}}$, resulting from the inclusion of a sub-dominant radiation bath, this being independent of $\Gamma/H_*$.
%%%%
}
%%%%
\label{efolds}
%%%%
\end{center}
%%%%
\end{figure}

Before discussing the resulting PBH mass function, let us calculate the time when the transition between production mechanisms occurs. In the simplest case, where $\Gamma$ is constant and $H^2\propto a^{-3}$ during reheating, PBH production becomes suppressed after
\be
\label{standard}
N_{\rm x} \simeq \frac{2}{3}{\rm ln}\left(\frac{2\sqrt{2}H_*\sigma}{3w_{\rm r}\Gamma}{\rm Erfc}^{-1}\left( \beta(\sigma)\right)
\right)
\ee
e-folds from the end of inflation, this indicating when the transition between production mechanisms occurs. Here $H_*\leq 8\times 10^{13}$ GeV is the scale of inflation \cite{Ade:2015lrj}. In the standard analysis of PBH formation during an early matter-dominated era, the total number of e-folds of reheating is
\begin{equation}\label{key}
N_{\rm reh} = \frac{2}{3}\,\mathrm{ln}\,\Big(\frac{H_*}{\Gamma}\Big) \,.
\end{equation}
Therefore, if we define $\Delta N \equiv N_{\rm reh} - N_{\rm x}$ as the change in the number of e-folds when one includes the effect of the subdominant radiation heat bath on PBH formation, Eqs.~(\ref{standard}) and (\ref{key}) imply
\begin{equation}
%\label{key}
\Delta N  = \frac{2}{3}\,\mathrm{ln}\,\Big(
%\frac{3w_r}{\sqrt{2}\sigma}
\frac{3w_r}{2\sqrt{2} \sigma \, \mathrm{Erfc}^{-1}
%(0.056\sigma^5)
\left( \beta(\sigma)\right)}
\Big) \,.
\end{equation}
As shown in Fig. \ref{efolds}, the reduction is $\mathcal{O}(10\%)$. It is also independent of $\Gamma$, because changing $\Gamma$ affects both when the suppression of PBH production occurs and when reheating is complete.

%%%%

\section{Primordial black hole mass function}
\label{sec:massfunction}

We define the current PBH mass function as
\be
\psi(M)\equiv  \frac{1}{\rho_{\rm DM}} \frac{{\rm d}\rho_{\rm PBH}(M)}{{\rm d}M}\,, 
\ee
where $\rho_{\rm DM}$ is the (current) dark matter density, so that the fraction of the DM density in PBHs in the mass interval $(M,M+{\rm d}M)$ is $\psi(M){\rm d}M$. During reheating, the ratio $\rho_{\rm PBH}/\rho_{\rm tot}$ remains constant and for PBHs forming in this period, we can identify the fraction of the Universe in PBHs at formation with the fraction at the reheating epoch. The current PBH DM fraction in the mass range $(M,M+{\rm d}M)$ is then
\bea \label{frelmd}
\psi(M){\rm d}M &= \frac{a_{\rm eq}}{a_{\rm reh}} \frac{\beta(M)}{M}\,  {\rm d}M \\
&= \left(\frac{g_*(T_{\rm reh})}{g_*(T_{\rm eq})}\right)^{1/3}\frac{T_{\rm reh}}{T_{\rm eq}} \frac{\beta(M)}{M} {\rm d}M \,,
\eea
where $g_*(T_{\rm eq})=3.909$ and $T_{\rm eq}=0.8$ eV \cite{Ade:2015xua}. 

When expressions~\eqref{kp} or \eqref{betamat} exceed expression~\eqref{betarad}, in order to solve for the resulting mass function, one has to specify the smallest and largest scales which become non-linear during the effective matter-dominated era, i.e. such that the expected perturbation $\sigma$ reaches unity by the time of the transition in the PBH production mechanism. The scales are given by
\be \label{masses}
M_{\rm min} = 4\pi\frac{M_{\rm P}^2}{H_*}, \quad
M_{\rm max} = 4\pi\sigma_{\rm max}^{3/2}\frac{M_{\rm P}^2}{H_{\rm RD}} \, ,
\ee
where $\sigma_{\rm max}$ is the variance of the density contrast at the time the scale $M_{\rm max}$ enters the horizon and $H_{\rm RD}=\Gamma$ for $\sigma\geq \sigma_c\simeq 0.05$, as discussed in Section \ref{reheating}. However, for $\sigma < \sigma_c$ we have 
\be
H_{\rm RD}
\equiv H_*e^{-3N_{\rm x}/2}
\simeq \frac{3w_{\rm r}}{2\sqrt{2}\sigma_{\rm max}}
\frac{\Gamma}{{\rm Erfc}^{-1}
\left(\beta(\sigma_{\rm max})\right)
}\,.
\ee
By using Eq.~\eqref{Treh}, we can write the scales as
\bea
M_{\rm min} &\simeq 7\times 10^{-34}\left(\frac{H_*}{10^{14} {\rm GeV}}\right)^{-1} M_\odot \,, \\
M_{\rm max} &\simeq 0.1\times
\begin{cases}
\sigma^{3/2}_{\rm max}\left(\frac{T_{\rm reh}}{\rm GeV}\right)^{-2}M_\odot \,,\\
\sigma^{5/2}_{\rm max}{\rm Erfc}^{-1}
\left(\beta(\sigma_{\rm max})\right)
\left(\frac{T_{\rm reh}}{\rm GeV}\right)^{-2}M_\odot ,
\end{cases}
\label{cutoff}
\eea
where the upper expression for $M_{\rm max}$ applies for $\sigma \geq \sigma_c$ and the lower for $\sigma < \sigma_c$.

\begin{figure}
%%%%
\begin{center}
%%%%
\includegraphics[width=.465\textwidth]{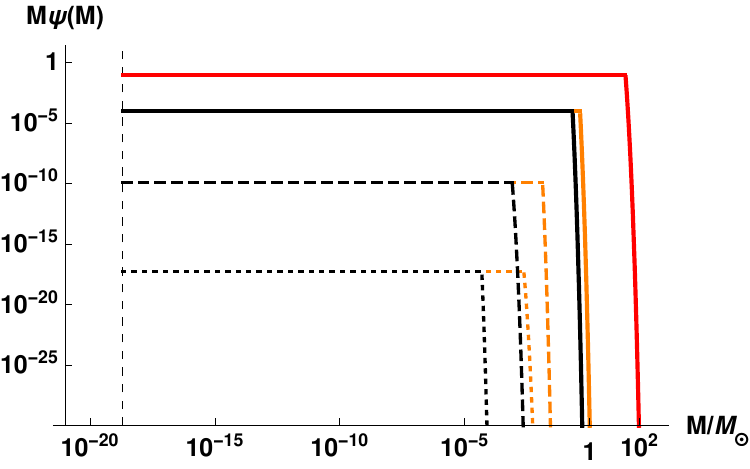}
%%%%
\includegraphics[width=.465\textwidth]{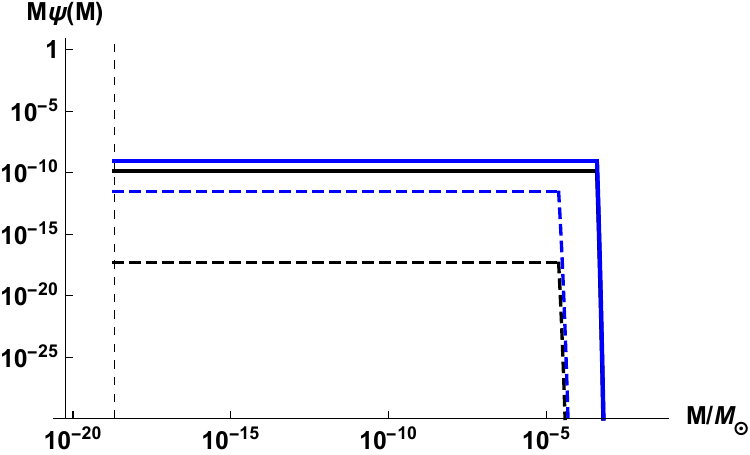}
%%%%
\caption{{\it Upper panel}: 
PBH mass function for $\sigma=10^{-2}$ (solid curves), $\sigma=10^{-3}$ (dashed curves) and $\sigma=10^{-3.5}$ (dotted curves). The black (orange) curves show the result with (without) the subdominant heat bath, assuming $T_{\rm reh}=0.01$ GeV. The red line corresponds to the critical case $\sigma=\sigma_c\simeq 0.05$, where the results with and without the subdominant heat bath coincide. In this figure $T_{\rm reh}=0.004$ GeV for this curve, giving the maximum PBH mass in this scenario. The lower cut-off is given by the mass of the PBHs evaporating at the present epoch, $M_{\rm evap} \simeq 2\times 10^{-19}M_\odot$.
%%%%
{\it Lower panel}: The PBH mass functions with (black curves) and without (blue curves) the effect of angular momentum for $\sigma=10^{-3}$ (solid curves) and $\sigma=10^{-3.5}$ (dashed curves). 
%%%%
We see that neglecting spin greatly overestimates the mass function for small $\sigma$.
%%%%
}
%%%%
\label{massfunction}
%%%%
\end{center}
%%%%
\end{figure}

In  the $\sigma < \sigma_c$ case, when the transition between production mechanisms occurs before reheating, the scales that enter the horizon after the last one to become non-linear, $M>M_{\rm max}$, are not expected to form PBHs. Note that $M_{\rm max}$ is considerably smaller than in the case where %$\beta_{\rm r}$
the usual $\beta$ for MD is taken to hold all the way to $\Gamma/H=1$. This and a few example mass functions are shown in Fig. \ref{massfunction} for scale-invariant $\sigma$. 

An important cosmologial question is whether PBHs can be large enough to explain the LIGO events. Since $M_{\rm max}$ increases with $\sigma$ and the maximum value of $\sigma$ allowed by our scenario is $\sigma_c\simeq 0.05$, this corresponds to a maximum cut-off at around $100\,M_{\odot}$. Although the PBH constraints would seem to preclude them providing all the dark matter \cite{Carr:2017jsz}, they might still be able to explain the LIGO events.
Our findings show that the effect of the subdominant radiation bath can be 
relevant for LIGO PBHs, provided $\Gamma$ is small enough.
A larger value of $\sigma$ would increase $M_{\rm max}$ up to $\mathcal{O}(10^3)M_\odot$ but in that case the subdominant heat bath would not affect the PBH formation rate.

Note that the mass function cuts off suddenly below the mass $M_{\rm min}$  given by Eq.~\eqref{cutoff}. However, if critical collapse can occur in a matter-dominated era (which is uncertain since this may require the presence of pressure), then one would expect the PBH mass function to have a low-mass tail below $M_{\rm min}$. In the radiation-dominated case this  tail has the form $\psi (M)  \propto M^{2.85}$, so its contribution  to the total PBH density is small, but it is not clear what it would be in the  $w \ll 1$ case. However, for $H_*\gtrsim 0.4$ GeV, that is for any sensible scale of inflation, the physical lower cut-off for the present-day mass function is given by the mass of the PBHs evaporating at the present epoch, $M_{\rm evap} \simeq 2\times 10^{-19}M_\odot$ \cite{Carr:2009jm}. This is included in Fig. \ref{massfunction}.

%%%%

The total dark matter fraction in PBHs is obtained by integrating the PBH mass function over mass:
\be
\label{totalenergy}
f =\frac{\rho_{\rm PBH}}{\rho_{\rm DM}} = \int_{M_{\rm min}}^{M_{\rm max}} \psi(M) {\rm d}M \,,
\ee
where the upper and lower limit of integration are given by Eq.~\eqref{masses}. Observational constraints on the final PBH abundance can then be evaluated using the method presented in Ref.~\cite{Carr:2017jsz}. In this paper our purpose is to
investigate how the subdominant heat bath truncates the PBH mass function, so we do not consider such constraints here.

A scenario where $\sigma$ is not scale-invariant can be considered once the primordial power spectrum is specified. The power spectrum is related to $\sigma(M)$ in a straightforward manner, as discussed recently in the context of PBHs in Refs. \cite{Carr:2017edp,Cole:2017gle,Kohri:2018qtx}. For example, the mass function skews towards the low mass end for a blue-tilted spectrum  and towards the high mass end for a red-tilted spectrum. We leave these aspects for future work.

%%%%%%%%%%%%%%%%%%%%%%%%%%%%%%%%%%%%%%%%%%%%%%%%%%%%%%%%%%%%%%%%%%%%%%%%%%%%%%%%%%%%%%%%%%

\section{Conclusions}
\label{conclusions}

In this paper we have studied the formation of PBHs in the early Universe during a period of slow reheating after inflation. We have calculated how the formation mechanism for PBHs makes a gradual transition during the change from the 
matter-dominated to radiation-dominated phases and quantified under what conditions this happens. 
We have found that the production mechanism never changes during reheating for the variance of the density contrast $\sigma\gtrsim \sigma_c\simeq 0.05$, so the Universe can be modelled as purely matter-dominated until the end of reheating, as in the standard approach. However, for $\sigma < \sigma_c$, the production mechanism smoothly changes even before reheating completes.
This happens when the ratio of the decay rate of the field responsible for reheating to the Hubble rate, $\Gamma/H$, takes the value given by Eq.~\eqref{GoverH}. However, we have found that the reduction in the number of e-folds after which the PBH production mechanism changes is independent of $\Gamma$, showing an $\mathcal{O}(10\%)$ change compared to the usual result.

We have calculated the PBH mass function and shown that for $\sigma<\sigma_c$ the maximum PBH mass can be considerably smaller than in the usual case, where PBH production continues until $\Gamma/H=1$. There is a maximum cut-off for the PBH mass at around $100\,M_{\odot}$, making the scenario particularly interesting for the recent LIGO observations of black hole mergers.
Although we have focussed on scenarios in which PBHs form during a slow-reheating period after inflation, we emphasize that the subdominant heat bath produced by a gradually decaying matter component is also likely to affect other scenarios, e.g. one where the matter-dominance is caused by metastable massive particles in general.
It would be interesting to see how this affects concrete models for reheating or other early Universe scenarios.

%%%%%%%%%%%%%%%%%%%%%%%%%%%%%%%%%%%%%%%%%%%%%%%%%%%%%%%%%%%%%%%%%%%%%%%%%%%%%%%%%%%%%%%%%%

%%%%%%%%%%%%%%%%%%%%%%%%%%%%%%%%%%%%%%%%%%%%%%%%%%%%%%%%
\section*{Acknowledgements}
K.D. is supported (in part) by the Lancaster-Manchester-Sheffield Consortium for Fundamental Physics under the STFC grant ST/L000520/1. C.O. is supported by the FST of Lancaster University. T.T. is supported by the STFC grant ST/J001546/1.

%%%%%%%%%%%%%%%%%%%%%%%%%%%%%%%%%%%%%%%%%%%%%%%%%%%%%%%%
\bibliography{PBH_RH.bib}

\end{document}